\documentclass[sigconf,screen]{acmart}

\usepackage{microtype}
\usepackage{balance}
\usepackage{xspace}
\usepackage{listings}
\usepackage{url}
\usepackage{hyperref}
\usepackage{float}

\AtBeginDocument{%
  \providecommand\BibTeX{{%
    \normalfont B\kern-0.5em{\scshape i\kern-0.25em b}\kern-0.8em\TeX}}}

\setcopyright{acmcopyright}
\acmPrice{15.00}
\acmDOI{10.1145/3395363.3404362}
\acmYear{2020}
\copyrightyear{2020}
\acmSubmissionID{issta20tool-id1-p}
\acmISBN{978-1-4503-8008-9/20/07}
\acmConference[ISSTA '20]{Proceedings of the 29th ACM SIGSOFT International Symposium on Software Testing and Analysis}{July 18--22, 2020}{Virtual Event, USA}
\acmBooktitle{Proceedings of the 29th ACM SIGSOFT International Symposium on Software Testing and Analysis (ISSTA '20), July 18--22, 2020, Virtual Event, USA}

\begin{document}

\newcommand{\gv}{G\&V\xspace}
\newcommand{\simsort}{$SimSort$\xspace}
\newcommand{\dist}{$dist$\xspace}
\newcommand{\secref}[1]{\S \ref{#1}}

\newcommand{\objsim}{ObjSim\xspace}
\newcommand{\objectutils}{\texttt{object-utils}\xspace}
\newcommand{\difftgen}{DiffTGen\xspace}
\newcommand{\opad}{OPad\xspace}
\newcommand{\fixtofit}{Fix2Fit\xspace}
\newcommand{\unsat}{UnsatGuided\xspace}
\newcommand{\qlose}{Qlose\xspace}
\newcommand{\ods}{ODS\xspace}

\newcommand{\prapr}{PraPR\xspace}

\definecolor{darkblue}{rgb}{0.0,0.0,0.6}
\lstdefinelanguage{XML}
{
  morestring=[b]",
  morestring=[s]{>}{<},
  morecomment=[s]{<?}{?>},
  stringstyle=\color{black},
  identifierstyle=\color{cyan},
  keywordstyle=\color{darkblue},
  showstringspaces=false,
  basicstyle={\small\ttfamily},
  morekeywords={artifactId,version,groupId,plugin,configuration,failingTests}%
}

\lstdefinelanguage{CSV}
{
  basicstyle={\small\ttfamily},
}

\title[\objsim: Lightweight Automatic Patch Prioritization \textit{via} Object Similarity]{\objsim: Lightweight Automatic Patch Prioritization\\\textit{via} Object Similarity}

\author{Ali Ghanbari}
\affiliation{%
  \institution{University of Texas at Dallas}
  \city{Richardson}
  \state{TX 75080}
  \country{USA}
}
\email{ali.ghanbari@utdallas.edu}

\begin{abstract}
In the context of test case based automatic program repair (APR), patches that pass all the test cases but fail to fix the bug are called \emph{overfitted} patches. Currently, patches generated by APR tools get inspected manually by the users to find and adopt genuine fixes. Being a laborious activity hindering widespread adoption of APR, automatic identification of overfitted patches has lately been the topic of active research. This paper presents engineering details of \objsim: a fully automatic, lightweight similarity-based patch prioritization tool for JVM-based languages. The tool works by comparing the system state at the exit point(s) of patched method before and after patching and prioritizing patches that result in state that is more similar to that of original, unpatched version on passing tests while less similar on failing ones. Our experiments with patches generated by the recent APR tool \prapr for fixable bugs from Defects4J v1.4.0 show that \objsim prioritizes 16.67\% more genuine fixes in top-1 place. A demo video of the tool is located at \url{https://bit.ly/2K8gnYV}.
\end{abstract}


\maketitle

\section{Introduction}\label{sec:introduction}
Manual debugging is notoriously difficult and costly. Automatic program repair (APR) \cite{apr2019cacm} aims to reduce the costs by suggesting high-quality patches that either directly fix the bugs or help the human developers during manual debugging. Generate-and-validate (\gv) refers to the class of APR techniques that attempt to fix the bug by first generating a pool of patches and validating the patches via certain rules and/or checks. A patch is said to be \textit{plausible} if it passes all the checks. Ideally, we would apply a sound method (e.g., formal verification) for checking validity of generated patches. However, in real-world situations, formal specifications of software are usually absent and due to theoretical limitations, formal verification is in general not automatable. In contrast, testing is the prevalent, economic method of getting more confidence about the quality of software. So, a vast majority of \gv APR techniques, indeed almost all APR techniques, use test cases as correctness criteria for patches \cite{bib:GMM17}.

A test-based \gv APR algorithm receives as input a buggy program and, optionally, a test suite consisting of at least one failing test identifying the bug, and produces zero or more plausible patches, i.e., patches that pass all the tests. A typical APR process in this class starts with fault localization to locate likely faulty program locations. It then attempts to fix the bug by applying a number of transformation operators on the identified suspicious locations to obtain a pool of candidate patches to be tested against the existing test suite. Patches that pass all the test cases are reported as plausible patches.

Unfortunately, test cases only partially specify the behavior of the programs and many of the generated patches happen to pass all of the test cases without actually fixing the bug. This makes APR techniques produce many plausible but incorrect patches, aka \emph{test case overfitted} patches \cite{overfitting15} (or simply overfitted patches). The process of examining APR-generated patches has to be manual for the oracle problem is undecidable, but in some cases, manually analyzing each and every one of the plausible patches could be even costlier than directly fixing the bug \cite{overfitting15}. Thus, a convenient method for post-processing the generated patches before reporting them to the developers is a need. This need is particularly pronounced in the case of APR techniques that are able to explore large search spaces and finding genuine patches that are reported after tens of incorrect ones (e.g., in \cite{ghanbari2019}).

This paper introduces \objsim, a lightweight, fully automatic patch prioritization tool based on object similarity heuristic. Users of test case based \gv APR techniques designed for JVM-based languages \cite{wiki:jvmLangs} are the envisioned users of this tool. \objsim prioritizes patches that are more likely to be correct so that the users of APR techniques spend less time examining the generated patches. The tool works by comparing system state at the exit point(s) of the patched method before and after patching and prioritizing patches that result in system state that is more similar to that of original, unpatched version on passing test cases while less similar on failing tests. The idea is that the behavior of the original program on passing test cases can be used as a partial specification of the desired behavior of the system, so we want a plausible patch to not only not to break passing test cases but also end up in a state similar to that of original version on passing tests. Meanwhile, we want the patch to deviate from original version on failing tests, thereby requiring patches to avoid known erroneous states.

We have applied \objsim on 358 plausible patches produced by the recent APR tool \prapr for 55 fixable bugs from Defects4J v1.4.0 \cite{newd4j}. We conducted our experiments on a commodity PC and set a time limit of 5 minutes and 16 GB of heap space for each bug. Compared to the original ranking scheme of \prapr, \objsim prioritizes 5 more genuine fixes in top-1 position (16.67\% improvement) and reduces average rank of genuine fixes from 3.04 to 2.74 (almost 10\% improvement), yet by relying only on runtime data, it is JVM language agnostic. \objsim, along with more details about our experiments, is publicly available \cite{objsimgithub}.


\section{A Top-Down Overview of \objsim}\label{sec:approach}
Virtually all available \gv APR techniques make one-point, or at most one-hunk, changes to the subject programs \cite{bib:GMM17}. Thus, we built \objsim based on the assumption that patching happens inside a single method. The basic idea of similarity-based patch prioritization is to compare system state at the exit point(s) of the patched method before and after patching and prioritizing patches that result in system state that is more similar to that of original, unpatched version on passing test cases while less similar on failing tests.
\begin{figure}
    \centering
    \includegraphics[scale=0.48]{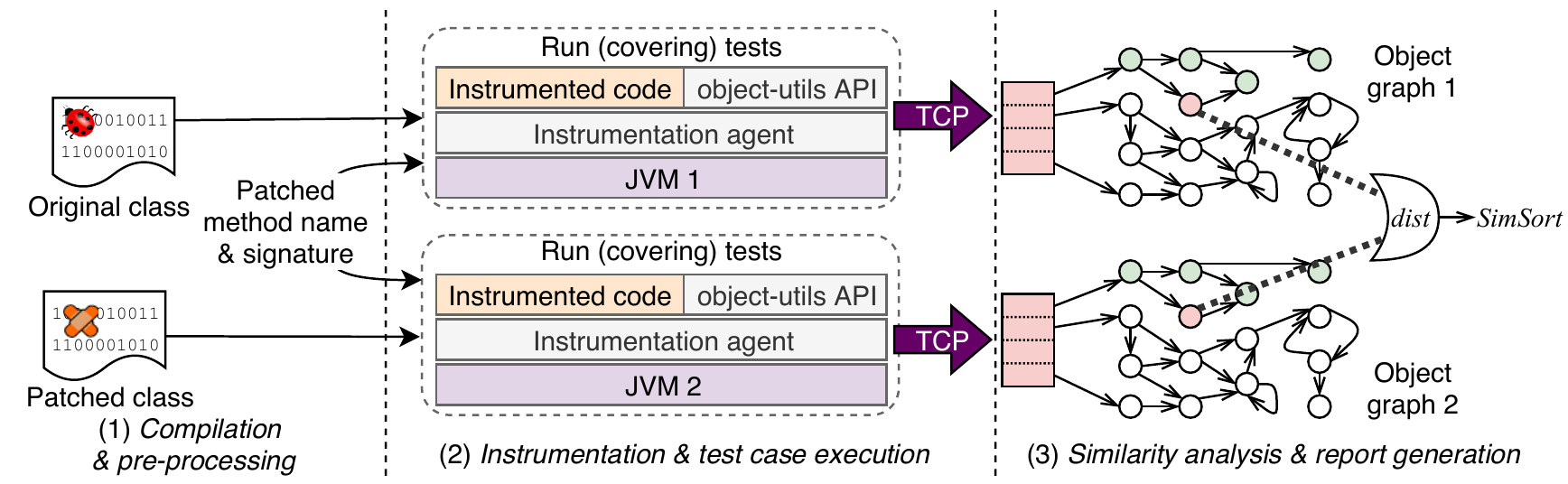}
    \caption{Overall workflow of \objsim, given class files for original and patched classes, full name of the patched method, and the set of test cases covering the patched method. DFS traversal of object graphs and distance calculator for computing similarity of objects used in \simsort is also depicted.}
    \label{fig:objsim}
\end{figure}

Figure \ref{fig:objsim} depicts an architectural overview of \objsim and the steps taken for calculating similarity of system state at the exit point(s) of patched method between original and patched versions. \objsim starts by obtaining class files for original and patched versions of the patched class and full name of the patched method. Some APR tools (e.g., \cite{ghanbari2019}) already provide the needed artifacts/information; in case any of these are absent (e.g., \cite{astor16} that produces only a source-level patch file), we may obtain them by applying the patch on the corresponding source file, compiling the file, and obtaining the name of the patched method via diffing.

The tool then instruments the original and patched versions of the patched method so that the instrumented program will capture a snapshot of the system state at every exit point of the patched method. This is done with the help of Java Agent technology \cite{oracle:javaAgent} and using Javassist library \cite{javassist} which allows inserting \emph{after advices} in the form of \texttt{finally} blocks. The instrumentation code uses Java reflection to capture and serialize the object graphs reachable from all the parameters of the patched method. Then separate processes (i.e., JVM 1 and JVM 2 in Figure \ref{fig:objsim}) are created to execute passing/failing tests against the instrumented programs while containing side-effects of test execution. Each version of instrumented program is once executed against (covering) originally passing test cases and once against (covering) originally failing tests. Note that restriction to only covering test cases is not necessary and it is done solely for speeding up the entire process. Note also that the two JVM instances can run in parallel to further speed up the prioritization process. Each round of test case execution results in a number of system state snapshots for original and patched version of the instrumented program.

We use $S(I,m,t)$ to denote the set of snapshots of system state at the exit point(s) of a method $m$ in the instrumented program $I$ resulting from executing covering test case $t$. Note that $|S(I,m,t)|\geq 1$, as a method might be called multiple times. Given a patch $\pi$ targeting method $m$ in the program $I$, \objsim obtains $S(I,m,t)$ and $S(\pi(I),m,t)$, where $\pi(I)$ denotes program $I$ with patch $\pi$ applied on it, for all tests $t$ of the program. Note further that these sets are constructed inside separate processes (namely JVM 1 or JVM 2 in Figure \ref{fig:objsim}), so we have to send them over to the parent process. \objsim establishes TCP connections between itself and the child processes and transfers the recorded sets by writing them on the socket. It is worth noting that serializing arbitrary Java objects is a non-trivial engineering undertaking and we omit its details here due to space limitations. We encourage the readers to visit the website of the companion library \objectutils for more information \cite{objectutils}.

Except in the case of patching operations that are regarded as \emph{anti-patterns} \cite{antipatterns2016}, and are usually avoided by modern APR techniques, patching a program does not change its control flow in significant ways. Thus, the number of times a method gets called tends to be the same before and after patching, i.e., $|S(I,m,t)|=|S(\pi(I),m,t)|$ for all tests $t$. In case the condition does not hold, \objsim puts the corresponding patch in a bucket $W$, which is to be ranked based on suspiciousness values (e.g., Ochiai suspiciousness) of the patched locations. For a given program $I$ with the set of test cases $t_1,\dots,t_n$, a patch $\pi\not\in W$ targeting method $m$, we use $l(\pi)$ to denote the sequence $\langle|S(\pi(I),m,t_1)|,\dots,|S(\pi(I),m,t_n)|\rangle$.

Let $P$ be the set of all plausible patches generated by the underlying APR tool. Let $\sim$ denote a binary relation over $P$ defined as $\pi_1\sim\pi_2$ iff $l(\pi_1)=l(\pi_2)$, where $\pi_1,\pi_2\not\in W$. It is easy to see that $\sim$ is an equivalence relation which is simply relating patches that have sequence of system state snapshots of the same length, both before and after patching. Let $\mathcal{Q}=P/\sim=\{\lbrack\pi\rbrack_\sim\mid\pi\in (P-W)\}$ be the quotient set of $P$ induced by $\sim$, which is simply the set of sets of patches that have the same value for $l$.

Having the sets $W$ and $\mathcal{Q}$, \objsim produces the final ranking by concatenating $m$ non-empty sequences $\sigma_1, \dots, \sigma_m$ where $\sigma_i$ is either $\langle\pi\rangle$ with $\pi\in W$ or $SimSort(Q)$ with $Q\in \mathcal{Q}$. $SimSort(Q)$ denotes a sequence of patches in $Q$ that is sorted according to similarity-based criteria. Clearly, $m=|W|+|\mathcal{Q}|$. The final sequence is sorted in such a way that $\sigma_i$ precedes $\sigma_j$ iff $MaxSusp(\sigma_i)\geq MaxSusp(\sigma_j)$, where $MaxSusp(\sigma)$ denotes the maximum suspiciousness value (e.g., Ochiai) for the patched locations corresponding to the patches in sequence $\sigma$. In what follows, we present a more detailed explanation on the algorithm for computing \simsort.

\subsection{\texorpdfstring{Computing \simsort}{Computing SimSort}}
Given a set $Q$ of patches, \simsort returns a sorted sequence of the patches in $Q$. Sorting is done by assigning a score to each patch in $Q$ and putting the patches according to their scores in descending order. In order to present the algorithm in a more precise way, we define distance matrix $D=[d_{ij}]_{|Q|\times n}$, where $n$ is the number of all test cases in the program. Each row in $D$ corresponds to a patch in $Q$ and each entry of the matrix represents the \emph{average distance} of system state in the patched program from that of the original program under some test $t$. Specifically, $d_{ij}=\mathrm{avg}\{dist(S_o,S_p)\mid S_o\in S(I,m,t_j)\wedge S_p\in S(\pi_i(I),m,t_j)\}$ where $t_j$ is a failing or passing test case. The function \dist computes distance between two objects. A more detailed description of this function is presented in the subsection that follows.

\simsort uses $D$ to compute scores matrix $R=[r_{ij}]_{|Q|\times n}$ as follows. For each $1\leq j\leq n$, the function sorts $j^\mathrm{th}$ column of $D$ in ascending (descending) order if $t_j$ is a passing (failing) test case. $r_{ij}$ is the position of $\pi_i$ in the sorted $j^\mathrm{th}$ column of $D$. The function obtains score of each patch $\pi_i$ by averaging $i^\mathrm{th}$ row of $R$. The final result returned by \simsort is obtained by sorting the patches based on their scores in reverse order.

\subsubsection{\texorpdfstring{Computing \dist}{Computing dist}}
Given two objects $s_1$ and $s_2$ (that could be system states), $dist(s_1,s_2)$ is computed via DFS traversal of the object graphs reachable from $s_1$ and $s_2$ and accumulating distances of ``sub-objects'' of the objects in a recursive manner. Specifically, \dist is defined recursively as follows.
\begin{itemize}
    \item $dist(s_1, s_2)=0$ if $s_1$, $s_2$ are equal references or equal primitive-typed objects.
    \item $dist(s_1, s_2)=1$ if $s_1$, $s_2$ are unequal primitive-typed objects of the same type.
    \item $dist(s_1, s_2)=\mathrm{Levenshtein~distance~between~}s_1\mathrm{~and~}s_2$, if $s_1$, $s_2$ are arrays of the same component type.
    \item $dist(s_1, s_2)=\sum_{i=1}^{n}dist(v(f_i, s_1),v(f_i, s_2))$ if $s_1$, $s_2$ are objects of the same type $\tau$ and $f_1,\dots,f_n$ are the names of the fields declared or inherited by $\tau$. Furthermore, $v(f,o)$ is defined to be the value of field $f$ for object $o$.
    \item $dist(s_1, s_2)=+\infty$ if $s_1$, $s_2$ are objects of different types.
\end{itemize}

For the sake of simplicity in presentation, we have assumed that the object graphs reachable from $s_1$, $s_2$ are acyclic. Many engineering details are also omitted. Please see our implementation \cite{objectutils} for more details. The rationale behind the above rules is to extend Levenshtein distance algorithm \cite{wiki:leven} to handle arbitrary objects: the distance between a primitive-typed value and another of the same type is 1, the distance between arrays is computed as per the conventional Levenshtein distance algorithm, object distances are computed field-wise in a recursive manner, and the distance between objects of different types is defined to be positive infinity.

\section{\objsim Usage}\label{sec:demo}
After checking out \objsim from \cite{objsimgithub} and installing it on the local Maven repository, the tool will be available in the form of a Maven plugin. In order to use \objsim to prioritize plausible patches, the user needs to add the following snippet under \texttt{<plugins>} tag in the POM file of the target project and list fully qualified names of the failing test cases under the designated tag.

\begin{lstlisting}[language=xml]
<plugin>
    <artifactId>objsim</artifactId>
    <groupId>edu.utdallas</groupId>
    <version>1.0-SNAPSHOT</version>
    <configuration>
        <failingTests> 
            <!-- list of failing tests -->
        </failingTests>
    </configuration>
</plugin>
\end{lstlisting}

The tool expects a CSV file, named \texttt{input-file.csv}, under the base directory of the project. The input file is intended to contain information about the patches. Each row of this file describes a patch and has to have the following format.
\begin{lstlisting}[language=csv]
Id,Susp,Method,Class-File,Covering-Tests
\end{lstlisting}

Where \texttt{Id} is a unique integer identifier of the patch corresponding to the line, \texttt{Susp} is the suspiciousness value for the patch location, \texttt{Method} is the fully qualified name of the patched method used during instrumentation, \texttt{Class-File} is the name of the compiled class file of patched class, and \texttt{Covering-Tests} is the space-separated list of test cases covering the patched location. Test case names should always be of the form \texttt{ClassName.MethodName} where \texttt{ClassName} is the fully qualified name of the test class. It is worth noting that we have shipped \objsim with a tool to construct the input CSV file from fix reports generated by \prapr.

After setting up \objsim, the tool can be invoked via the command \texttt{mvn edu.utdallas:objsim:validate}. The output of the tool shall be a sorted list of patch identifiers stored in a text file under the base directory of the target project. For more information and a demo, please see the companion video at \url{https://bit.ly/2K8gnYV}.

\section{Related Work}\label{sec:related}
Automated patch classification has lately attracted the attention of APR research community \cite{ye2019automateda,xiong2018,opad2017,overfitting15,difftgen,unsatguided2018}. \objsim is most related to \difftgen \cite{difftgen} and the technique introduced in \cite{xiong2018}. \difftgen identifies overfitted patches by finding semantic differences between the original, buggy program and its patched version by presenting values of variables and fields to the user and asking them to decide if the demonstrated behavior is reasonable. On the other hand, \cite{xiong2018} is fully automatic and works by comparing execution traces between the original and patched programs near the patched method. Although there is a recent study using \difftgen for classifying patches \cite{ye2019automatedb}, it is still unclear whether or not asking the users to decide if a behavior is desired (esp., by printing out the intermediate results of computations) is cost-effective. Also, \cite{xiong2018} records details about program execution that might be unnecessary when we reason about final results of computations; not to mention that despite discarding information from recorded traces, the implementation still calls for a large amount of computational resources. Unlike \difftgen, \objsim automates the process by computing the similarity between system state at the exit point(s) of the patched method in the original program and its patched version. And unlike \cite{xiong2018}, it is lightweight yet it does not need to dismiss information about program behavior.

A body of research is also dedicated to techniques for repairing programs while minimizing semantic difference between the original and the patched versions. Chandra et al. \cite{chandra2011angelic} introduce a technique for identification of expressions in a buggy program that if replaced with a \emph{good} repair candidate, will solve the bug. A good repair candidate is the one that corrects the failing executions, yet does not break passing tests. This idea forms the basis of the technique for repairing reactive programs by taking a buggy program as a partial specification of the desired behavior and producing high-quality repairs by deviating from it as least as possible \cite{von2015program}. This is related to \qlose \cite{d2016qlose}, a technique for synthesizing fixes for small-scale student programs to pass all the test cases while the difference between execution traces of the original and the patched versions remains minimal. Although these works are related to \objsim in the basic idea of comparing runtime state of a given patch with that of original version, the goal of two lines of research is fundamentally different. While \objsim strives for achieving scalability in patch prioritization and applicability to a wide range of APR techniques \cite{astor16,wen18,chen2019sequencer} and programming languages, \qlose and \cite{von2015program} aim to synthesize a patch that is correct by construction and neither scalability nor applicability are concerns.

Pattern-based repair techniques \cite{le2016history,ghanbari2019,tbar19} generate patches based on the transformation operators learned from real-world bug fix commits with the goal of generating patches that are more likely to be correct. Anti-patterns \cite{antipatterns2016} refer to the transformation operators that commonly lead to plausible but incorrect patches. Similar pattern-based patch prioritization is employed by \cite{ghanbari2019,wen18}. \ods \cite{ye2019automateda} uses source code level features to discriminate correct patches from incorrect ones. Unlike these techniques, \objsim does not depend on program text, so it is JVM language agnostic and can be used to prioritize patches generated for programs written in languages other than Java, e.g., Kotlin or Scala.

We conclude this section by discussing other techniques. In \cite{opad2017}, Yang et al. introduce \opad that automatically filters out overfitted patches introducing regression by generating test cases so as to fuzz test the patched method and identify patches that manifest predefined erroneous behavior (e.g., memory leak or crash). Recently, Gao et al. introduce \fixtofit \cite{crashavoiding} that follows a similar approach. These techniques are not expected to be effective in case of programs written in managed programming languages \cite{xiong2018}. Le et al. \cite{semantic2018} show that semantic-based APR techniques also suffer from overfitting.

\section{Conclusions}\label{sec:conc}
This paper presents \objsim, a fully automatic, lightweight similarity-based patch prioritization tool for JVM-based languages. It works by comparing the system state at exit point(s) of the patched method between original program and its patched version, and prioritizing patches that result in system state that is more similar to that of original version on passing tests while less similar on failing tests. The key to scalability of \objsim is to record and compare computed object graphs rather than complete execution traces. We observed that this technique can be quite effective, resulting in 16.67\% improvement compared to default ranking scheme of state-of-the-art \prapr, yet, being semantic-based, the technique is language agnostic.
\balance
\section*{Acknowledgements}
The author thanks ISSTA reviewers for their insightful comments.
\balance
\bibliographystyle{ACM-Reference-Format}
\bibliography{bibdb}


\begin{thebibliography}{28}


\ifx \showCODEN    \undefined \def \showCODEN     #1{\unskip}     \fi
\ifx \showDOI      \undefined \def \showDOI       #1{#1}\fi
\ifx \showISBNx    \undefined \def \showISBNx     #1{\unskip}     \fi
\ifx \showISBNxiii \undefined \def \showISBNxiii  #1{\unskip}     \fi
\ifx \showISSN     \undefined \def \showISSN      #1{\unskip}     \fi
\ifx \showLCCN     \undefined \def \showLCCN      #1{\unskip}     \fi
\ifx \shownote     \undefined \def \shownote      #1{#1}          \fi
\ifx \showarticletitle \undefined \def \showarticletitle #1{#1}   \fi
\ifx \showURL      \undefined \def \showURL       {\relax}        \fi
\providecommand\bibfield[2]{#2}
\providecommand\bibinfo[2]{#2}
\providecommand\natexlab[1]{#1}
\providecommand\showeprint[2][]{arXiv:#2}

\bibitem[\protect\citeauthoryear{Chandra, Torlak, Barman, and Bodik}{Chandra
  et~al\mbox{.}}{2011}]%
        {chandra2011angelic}
\bibfield{author}{\bibinfo{person}{Satish Chandra}, \bibinfo{person}{Emina
  Torlak}, \bibinfo{person}{Shaon Barman}, {and} \bibinfo{person}{Rastislav
  Bodik}.} \bibinfo{year}{2011}\natexlab{}.
\newblock \showarticletitle{Angelic debugging}. In
  \bibinfo{booktitle}{\emph{ICSE}}. \bibinfo{pages}{121--130}.
\newblock


\bibitem[\protect\citeauthoryear{Chen, Kommrusch, Tufano, Pouchet, Poshyvanyk,
  and Monperrus}{Chen et~al\mbox{.}}{2019}]%
        {chen2019sequencer}
\bibfield{author}{\bibinfo{person}{Zimin Chen}, \bibinfo{person}{Steve~James
  Kommrusch}, \bibinfo{person}{Michele Tufano}, \bibinfo{person}{Louis-No{\"e}l
  Pouchet}, \bibinfo{person}{Denys Poshyvanyk}, {and} \bibinfo{person}{Martin
  Monperrus}.} \bibinfo{year}{2019}\natexlab{}.
\newblock \showarticletitle{Sequencer: Sequence-to-sequence learning for
  end-to-end program repair}.
\newblock \bibinfo{journal}{\emph{TSE}} (\bibinfo{year}{2019}).
\newblock


\bibitem[\protect\citeauthoryear{Chiba}{Chiba}{2000}]%
        {javassist}
\bibfield{author}{\bibinfo{person}{Shigeru Chiba}.}
  \bibinfo{year}{2000}\natexlab{}.
\newblock \bibinfo{howpublished}{\url{https://bit.ly/2UmMuIT}}.
\newblock
\newblock
\shownote{Accessed: April 2020.}


\bibitem[\protect\citeauthoryear{D’Antoni, Samanta, and Singh}{D’Antoni
  et~al\mbox{.}}{2016}]%
        {d2016qlose}
\bibfield{author}{\bibinfo{person}{Loris D’Antoni}, \bibinfo{person}{Roopsha
  Samanta}, {and} \bibinfo{person}{Rishabh Singh}.}
  \bibinfo{year}{2016}\natexlab{}.
\newblock \showarticletitle{Qlose: Program repair with quantitative
  objectives}. In \bibinfo{booktitle}{\emph{CAV}}. \bibinfo{pages}{383--401}.
\newblock


\bibitem[\protect\citeauthoryear{Gao, Mechtaev, and Roychoudhury}{Gao
  et~al\mbox{.}}{2019}]%
        {crashavoiding}
\bibfield{author}{\bibinfo{person}{Xiang Gao}, \bibinfo{person}{Sergey
  Mechtaev}, {and} \bibinfo{person}{Abhik Roychoudhury}.}
  \bibinfo{year}{2019}\natexlab{}.
\newblock \showarticletitle{Crash-avoiding Program Repair}. In
  \bibinfo{booktitle}{\emph{ISSTA}}. \bibinfo{pages}{8--18}.
\newblock


\bibitem[\protect\citeauthoryear{Gay}{Gay}{2017}]%
        {newd4j}
\bibfield{author}{\bibinfo{person}{Gregory Gay}.}
  \bibinfo{year}{2017}\natexlab{}.
\newblock \bibinfo{howpublished}{\url{http://bit.ly/2vxSQwR}}.
\newblock
\newblock
\shownote{Accessed: April 2020.}


\bibitem[\protect\citeauthoryear{Gazzola, Micucci, and Mariani}{Gazzola
  et~al\mbox{.}}{2017}]%
        {bib:GMM17}
\bibfield{author}{\bibinfo{person}{Luca Gazzola}, \bibinfo{person}{Daniela
  Micucci}, {and} \bibinfo{person}{Leonardo Mariani}.}
  \bibinfo{year}{2017}\natexlab{}.
\newblock \showarticletitle{Automatic software repair: A survey}.
\newblock \bibinfo{journal}{\emph{TSE}} (\bibinfo{year}{2017}),
  \bibinfo{pages}{34--67}.
\newblock


\bibitem[\protect\citeauthoryear{Ghanbari}{Ghanbari}{2020a}]%
        {objsimgithub}
\bibfield{author}{\bibinfo{person}{Ali Ghanbari}.}
  \bibinfo{year}{2020}\natexlab{a}.
\newblock \bibinfo{howpublished}{\url{http://bit.ly/2I3aMBU}}.
\newblock
\newblock
\shownote{Accessed: April 2020.}


\bibitem[\protect\citeauthoryear{Ghanbari}{Ghanbari}{2020b}]%
        {objectutils}
\bibfield{author}{\bibinfo{person}{Ali Ghanbari}.}
  \bibinfo{year}{2020}\natexlab{b}.
\newblock \bibinfo{howpublished}{\url{https://bit.ly/2U4SUxt}}.
\newblock
\newblock
\shownote{Accessed: April 2020.}


\bibitem[\protect\citeauthoryear{Ghanbari, Benton, and Zhang}{Ghanbari
  et~al\mbox{.}}{2019}]%
        {ghanbari2019}
\bibfield{author}{\bibinfo{person}{Ali Ghanbari}, \bibinfo{person}{Samuel
  Benton}, {and} \bibinfo{person}{Lingming Zhang}.}
  \bibinfo{year}{2019}\natexlab{}.
\newblock \showarticletitle{Practical Program Repair via Bytecode Mutation}. In
  \bibinfo{booktitle}{\emph{ISSTA}}. \bibinfo{pages}{19--30}.
\newblock


\bibitem[\protect\citeauthoryear{Goues, Pradel, and Roychoudhury}{Goues
  et~al\mbox{.}}{2019}]%
        {apr2019cacm}
\bibfield{author}{\bibinfo{person}{Claire~Le Goues}, \bibinfo{person}{Michael
  Pradel}, {and} \bibinfo{person}{Abhik Roychoudhury}.}
  \bibinfo{year}{2019}\natexlab{}.
\newblock \showarticletitle{Automated Program Repair}.
\newblock \bibinfo{journal}{\emph{CACM}} (\bibinfo{year}{2019}),
  \bibinfo{pages}{56--65}.
\newblock


\bibitem[\protect\citeauthoryear{Le, Lo, and Le~Goues}{Le
  et~al\mbox{.}}{2016}]%
        {le2016history}
\bibfield{author}{\bibinfo{person}{Xuan Bach~D Le}, \bibinfo{person}{David Lo},
  {and} \bibinfo{person}{Claire Le~Goues}.} \bibinfo{year}{2016}\natexlab{}.
\newblock \showarticletitle{History driven program repair}. In
  \bibinfo{booktitle}{\emph{SANER}}. \bibinfo{pages}{213--224}.
\newblock


\bibitem[\protect\citeauthoryear{Le, Thung, Lo, and Le~Goues}{Le
  et~al\mbox{.}}{2018}]%
        {semantic2018}
\bibfield{author}{\bibinfo{person}{Xuan Bach~D Le}, \bibinfo{person}{Ferdian
  Thung}, \bibinfo{person}{David Lo}, {and} \bibinfo{person}{Claire Le~Goues}.}
  \bibinfo{year}{2018}\natexlab{}.
\newblock \showarticletitle{Overfitting in semantics-based automated program
  repair}.
\newblock \bibinfo{journal}{\emph{ESE}} (\bibinfo{year}{2018}),
  \bibinfo{pages}{3007--3033}.
\newblock


\bibitem[\protect\citeauthoryear{Liu, Koyuncu, Kim, and Bissyand\'{e}}{Liu
  et~al\mbox{.}}{2019}]%
        {tbar19}
\bibfield{author}{\bibinfo{person}{Kui Liu}, \bibinfo{person}{Anil Koyuncu},
  \bibinfo{person}{Dongsun Kim}, {and} \bibinfo{person}{Tegawend\'{e}~F.
  Bissyand\'{e}}.} \bibinfo{year}{2019}\natexlab{}.
\newblock \showarticletitle{TBar: Revisiting Template-Based Automated Program
  Repair}. In \bibinfo{booktitle}{\emph{ISSTA}}. \bibinfo{pages}{31--42}.
\newblock


\bibitem[\protect\citeauthoryear{Martinez and Monperrus}{Martinez and
  Monperrus}{2016}]%
        {astor16}
\bibfield{author}{\bibinfo{person}{Matias Martinez} {and}
  \bibinfo{person}{Martin Monperrus}.} \bibinfo{year}{2016}\natexlab{}.
\newblock \showarticletitle{ASTOR: A Program Repair Library for Java (Demo)}.
  In \bibinfo{booktitle}{\emph{ISSTA}}. \bibinfo{pages}{441--444}.
\newblock


\bibitem[\protect\citeauthoryear{{Oracle Corporation}}{{Oracle
  Corporation}}{2020}]%
        {oracle:javaAgent}
\bibfield{author}{\bibinfo{person}{{Oracle Corporation}}.}
  \bibinfo{year}{2020}\natexlab{}.
\newblock \bibinfo{title}{Java Agent}.
\newblock
\newblock
\urldef\tempurl%
\url{https://bit.ly/3czmzFV}
\showURL{%
\tempurl}
\newblock
\shownote{Accessed June, 2020.}


\bibitem[\protect\citeauthoryear{Smith, Barr, Le~Goues, and Brun}{Smith
  et~al\mbox{.}}{2015}]%
        {overfitting15}
\bibfield{author}{\bibinfo{person}{Edward~K Smith}, \bibinfo{person}{Earl~T
  Barr}, \bibinfo{person}{Claire Le~Goues}, {and} \bibinfo{person}{Yuriy
  Brun}.} \bibinfo{year}{2015}\natexlab{}.
\newblock \showarticletitle{Is the cure worse than the disease? overfitting in
  automated program repair}. In \bibinfo{booktitle}{\emph{FSE}}.
  \bibinfo{pages}{532--543}.
\newblock


\bibitem[\protect\citeauthoryear{Tan, Yoshida, Prasad, and Roychoudhury}{Tan
  et~al\mbox{.}}{2016}]%
        {antipatterns2016}
\bibfield{author}{\bibinfo{person}{Shin~H. Tan}, \bibinfo{person}{Hiroaki
  Yoshida}, \bibinfo{person}{Mukul~R Prasad}, {and} \bibinfo{person}{Abhik
  Roychoudhury}.} \bibinfo{year}{2016}\natexlab{}.
\newblock \showarticletitle{Anti-patterns in search-based program repair}. In
  \bibinfo{booktitle}{\emph{FSE}}. \bibinfo{pages}{727--738}.
\newblock


\bibitem[\protect\citeauthoryear{von Essen and Jobstmann}{von Essen and
  Jobstmann}{2015}]%
        {von2015program}
\bibfield{author}{\bibinfo{person}{Christian von Essen} {and}
  \bibinfo{person}{Barbara Jobstmann}.} \bibinfo{year}{2015}\natexlab{}.
\newblock \showarticletitle{Program repair without regret}. In
  \bibinfo{booktitle}{\emph{CAV}}. \bibinfo{pages}{26--50}.
\newblock


\bibitem[\protect\citeauthoryear{Wen, Chen, Wu, Hao, and Cheung}{Wen
  et~al\mbox{.}}{2018}]%
        {wen18}
\bibfield{author}{\bibinfo{person}{Ming Wen}, \bibinfo{person}{Junjie Chen},
  \bibinfo{person}{Rongxin Wu}, \bibinfo{person}{Dan Hao}, {and}
  \bibinfo{person}{Shing-Chi Cheung}.} \bibinfo{year}{2018}\natexlab{}.
\newblock \showarticletitle{Context-aware patch generation for better automated
  program repair}. In \bibinfo{booktitle}{\emph{ICSE}}. \bibinfo{pages}{1--11}.
\newblock


\bibitem[\protect\citeauthoryear{{Wikipedia contributors}}{{Wikipedia
  contributors}}{2020a}]%
        {wiki:leven}
\bibfield{author}{\bibinfo{person}{{Wikipedia contributors}}.}
  \bibinfo{year}{2020}\natexlab{a}.
\newblock \bibinfo{title}{Damerau–Levenshtein distance --- {Wikipedia}{,} The
  Free Encyclopedia}.
\newblock
\newblock
\urldef\tempurl%
\url{https://bit.ly/2BrMOAj}
\showURL{%
\tempurl}
\newblock
\shownote{Accessed June 2020].}


\bibitem[\protect\citeauthoryear{{Wikipedia contributors}}{{Wikipedia
  contributors}}{2020b}]%
        {wiki:jvmLangs}
\bibfield{author}{\bibinfo{person}{{Wikipedia contributors}}.}
  \bibinfo{year}{2020}\natexlab{b}.
\newblock \bibinfo{title}{List of JVM languages --- {Wikipedia}{,} The Free
  Encyclopedia}.
\newblock
\newblock
\urldef\tempurl%
\url{https://bit.ly/3714hvf}
\showURL{%
\tempurl}
\newblock
\shownote{Accessed June, 2020.}


\bibitem[\protect\citeauthoryear{Xin and Reiss}{Xin and Reiss}{2017}]%
        {difftgen}
\bibfield{author}{\bibinfo{person}{Qi Xin} {and} \bibinfo{person}{Steven~P
  Reiss}.} \bibinfo{year}{2017}\natexlab{}.
\newblock \showarticletitle{Identifying test-suite-overfitted patches through
  test case generation}. In \bibinfo{booktitle}{\emph{ISSTA}}.
  \bibinfo{pages}{226--236}.
\newblock


\bibitem[\protect\citeauthoryear{Xiong, Liu, Zeng, Zhang, and Huang}{Xiong
  et~al\mbox{.}}{2018}]%
        {xiong2018}
\bibfield{author}{\bibinfo{person}{Yingfei Xiong}, \bibinfo{person}{Xinyuan
  Liu}, \bibinfo{person}{Muhan Zeng}, \bibinfo{person}{Lu Zhang}, {and}
  \bibinfo{person}{Gang Huang}.} \bibinfo{year}{2018}\natexlab{}.
\newblock \showarticletitle{Identifying patch correctness in test-based program
  repair}. In \bibinfo{booktitle}{\emph{ICSE}}. \bibinfo{pages}{789--799}.
\newblock


\bibitem[\protect\citeauthoryear{Yang, Zhikhartsev, Liu, and Tan}{Yang
  et~al\mbox{.}}{2017}]%
        {opad2017}
\bibfield{author}{\bibinfo{person}{Jinqiu Yang}, \bibinfo{person}{Alexey
  Zhikhartsev}, \bibinfo{person}{Yuefei Liu}, {and} \bibinfo{person}{Lin Tan}.}
  \bibinfo{year}{2017}\natexlab{}.
\newblock \showarticletitle{Better test cases for better automated program
  repair}. In \bibinfo{booktitle}{\emph{FSE}}. \bibinfo{pages}{831--841}.
\newblock


\bibitem[\protect\citeauthoryear{Ye, Gu, Martinez, Durieux, and Monperrus}{Ye
  et~al\mbox{.}}{2019a}]%
        {ye2019automateda}
\bibfield{author}{\bibinfo{person}{He Ye}, \bibinfo{person}{Jian Gu},
  \bibinfo{person}{Matias Martinez}, \bibinfo{person}{Thomas Durieux}, {and}
  \bibinfo{person}{Martin Monperrus}.} \bibinfo{year}{2019}\natexlab{a}.
\newblock \showarticletitle{Automated Classification of Overfitting Patches
  with Statically Extracted Code Features}.
\newblock \bibinfo{journal}{\emph{arXiv}} (\bibinfo{year}{2019}).
\newblock


\bibitem[\protect\citeauthoryear{Ye, Martinez, and Monperrus}{Ye
  et~al\mbox{.}}{2019b}]%
        {ye2019automatedb}
\bibfield{author}{\bibinfo{person}{He Ye}, \bibinfo{person}{Matias Martinez},
  {and} \bibinfo{person}{Martin Monperrus}.} \bibinfo{year}{2019}\natexlab{b}.
\newblock \showarticletitle{Automated Patch Assessment for Program Repair at
  Scale}.
\newblock \bibinfo{journal}{\emph{arXiv}} (\bibinfo{year}{2019}).
\newblock


\bibitem[\protect\citeauthoryear{Yu, Martinez, Danglot, Durieux, and
  Monperrus}{Yu et~al\mbox{.}}{2018}]%
        {unsatguided2018}
\bibfield{author}{\bibinfo{person}{Zhongxing Yu}, \bibinfo{person}{Matias
  Martinez}, \bibinfo{person}{Benjamin Danglot}, \bibinfo{person}{Thomas
  Durieux}, {and} \bibinfo{person}{Martin Monperrus}.}
  \bibinfo{year}{2018}\natexlab{}.
\newblock \showarticletitle{Alleviating patch overfitting with automatic test
  generation: a study of feasibility and effectiveness for the Nopol repair
  system}.
\newblock \bibinfo{journal}{\emph{ESE}} (\bibinfo{year}{2018}),
  \bibinfo{pages}{33--67}.
\newblock


\end{thebibliography}
\end{document}